\documentclass[twocolumn,pra,superscriptaddress,showpacs]{revtex4}
\usepackage{graphicx,subfigure,amsmath,amssymb,hyperref,mathrsfs}
\usepackage{setspace}

\usepackage{color}
\newcommand{\op}[1]{\hat{#1}}

\newcommand{\ket}[1]{|{#1}\rangle}
\newcommand{\bn}{\begin{eqnarray}}
\newcommand{\en}{\end{eqnarray}}
\newcommand{\n}{\nonumber}
\newcommand{\h}{\hspace}

\begin{document}

\title{Criticality and Spin Squeezing in the Rotational Dynamics of a BEC on a Ring Lattice}

\author{M. Kol\'a\v r}
\affiliation{Department of Optics, Palack\'{y}
University, 771 46 Olomouc, Czech Republic}
\author{T. Opatrn\'y}
\affiliation{Department of Optics, Palack\'{y}
University, 771 46 Olomouc, Czech Republic}
\author{Kunal K. Das}
\affiliation{Department of Physical Sciences, Kutztown University of Pennsylvania, Kutztown, Pennsylvania 19530, USA}
\affiliation{Kavli Institute for Theoretical Physics, UCSB, Santa Barbara, CA 93106}

\begin{abstract}
We examine the dynamics of circulating modes of a Bose-Einstein condensate
confined in toroidal lattice. Nonlinearity due to interactions leads to criticality that separates oscillatory and self-trapped phases among counter-propagating modes which however share the same physical space. In the mean-field limit, the criticality is found to substantially enhance sensitivity to rotation of the system. Analysis of the quantum dynamics reveals the fluctuations near criticality are significant, that we explain using spin-squeezing formalism visualized on a Bloch sphere. We utilize the squeezing to propose a Ramsey interferometric scheme that suppresses fluctuation in the relevant quadrature sensitive to rotation.
\end{abstract}
\date{\today }
\maketitle

\section{Introduction} Lattice potentials have become an indispensable ingredient of the physics of ultracold atoms with their remarkable
impact already charted in several excellent reviews \cite{optical-lattices-RMP, Bloch-RMP-Many-Body, Dalibard-RMP-Artificial-Gauge}. More recently, there has been another exciting development in the study of cold atoms with their confinement in multiply-connected topology in the form of toroidal traps realized by a variety of techniques \cite{Boshier-painted-potential,Phillips-2007,ramanathan,Hadzibabic-spinor,Stamper-Kurn-2015}. Several physical phenomena that rely upon such geometry are being actively studied, including generation of persistent currents \cite{Phillips-2007}, and Superconducting Quantum Interference Device (SQUID) \cite{Boshier-SQUID}.  Even such phenomena that do not intrinsically require such topology are revealing novel features in a torus geometry, including atomic versions of Josehpson effect \cite{Wright-phase-slips,Solenov-Josephson}, Schr\"odinger cat  states \cite{Maria-Ray-macroscopic-superpositions}, Tonks-Girardeau gas \cite{Brand-tonks,Minguzzi-tonks}, dipolar \cite{Jezek-dipolar} and spinor condensates \cite{Hadzibabic-spinor} and vortex  \cite{Stringari-vortex,Brand-joshephson-vortex} and soliton dynamics \cite{Kavoulakis-soliton,Reatto-soliton}.

Considering the rich physics associated with cold atoms in optical lattices and in a torus topology separately, it is to be expected that a marriage of these two different kinds of periodicity will be at least equally rich \cite{Cataliotti-ring-lattice}. So far, there has only been a handful of studies of such a system. The focus has been on the influence of a lattice on some of the phenomena mentioned above \cite{ring-lattice-dipolar,ring-lattice-persistent-current,Tiesinga-soliton} and on the many body physics associated with the Bose-Hubbard model \cite{Piza-ring-lattice-superfluidity,Moreno-Bose-Hubbard,Jezek-Bose-Hubbard} that has been studied extensively in the context of optical lattices. However, Bose-Einstein condensate (BEC) in a ring-shaped lattice offer an unique combination of features: a natural rotation axis, dual periodicity and nonlinearity. Our goal in this paper is to examine the dynamics that results from the juxtaposition of all those features.

The circular topology of a ring trap makes it inevitable to consider the influence of and sensitivity to rotation of this system.  Specifically we find that the nonlinear behavior arising from inter-atomic interactions, when combined with the lattice potential leads to critical behavior that displays strong sensitivity to rotation due to the toroidal geometry. Interactions among the atoms also lead to spin-squeezing effects \cite{Kitagawa} that are particularly susceptible to both rotations and the criticality.  The goal of this paper is to examine all of these effects in tandem, using several different approaches. Particularly, we show that the dynamics can be used to implement a squeezed Ramsey interferometric scheme for rotation sensing.

We develop our physical model in Sec. II, and then use a mean field approximation in Sec. III to determine the basic dynamical features and explain the nature and origin of the critical behavior. We then go beyond the mean field picture in Sec. IV and study the quantum dynamics to probe the region around criticality to understand the significant impact of fluctuations. We confirm the validity in the limit of large particle numbers by using a Gaussian approximation for the quantum correlations. Having also shown in the earlier sections that the essential quantum dynamics can be mapped to a spin-squeezing Hamiltonian visualized on a Bloch sphere, in Sec. V, we use that to propose a Ramsey interferometric scheme for the purpose of reducing the increased fluctuations observed near criticality to improve rotation sensitivity. In Sec. VI we provide numerical estimates for the sensitivity to rotation based on physical parameters for the different scenarios considered. We conclude with a discussion of feasibility and challenges towards implementing our ideas in experiments and the prospects of practical utility.

\begin{figure}[t]
\centering
\includegraphics[width=\columnwidth]{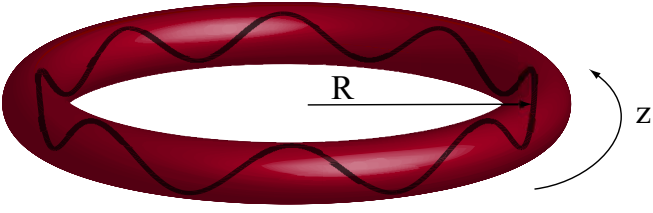}
\caption{(Color online) The atoms are trapped in a toroidal geometry with effectively one-dimensional dynamics about the major circle. A circular lattice potential along the ring can be introduced or removed as required.}
\label{Figure-1}
\end{figure}

\section{Physical Model}

We consider a BEC in a toroidal trap as shown in Fig.~\ref{Figure-1}. We take the minor radius to be much smaller than the major radius so that the system can be treated as a cylinder ${\bf r}=(z,r,\phi)$ with periodic boundary condition on $z$.  We assume the confinement along $(r,\phi)$, transverse to the ring circumference to be sufficiently strong to keep the atoms in the ground state
$\psi_r(r)\psi_{\phi}(\phi)$ for those degrees of freedom, so that the three-dimensional bosonic field operator can be written in the effective form $ \hat{\Psi}(z)\psi_r(r)\psi_{\phi}(\phi)$.
Integrating out the transverse degrees of freedom, the dynamics can be described by an effective one dimensional Hamiltonian
\begin{eqnarray}
&&\op{H}=\int_0^{2\pi R}{\rm d}z\op{\Psi}^\dagger(z)\times\\
&&\left[- \frac{\hbar^2}{2m}\partial^2_z+U(z,t)+\frac{g}{4\pi l_r^2} \op{\Psi}^\dagger(z)\op{\Psi}(z)\right]\op{\Psi}(z).\n
\label{QF-Hamiltonian}
\end{eqnarray}
where  $g=4\pi\hbar^2a/m$ is the interaction strength defined by the $s$-wave scattering length $a$, and $l_r$ is the average harmonic oscillator length for the transverse confinement. The potential along the ring is taken to be a periodic lattice, rotating with frequency $\omega$,
\begin{eqnarray}
U(z,t)=\hbar u_x \cos\left[2q ({\textstyle\frac{z}{R}} - \omega t) \right]+ \hbar u_y \sin\left[2q ({\textstyle\frac{z}{R}} - \omega t) \right] .
\end{eqnarray}
Expanding the field operator in the eigenstates of the ring
\bn \op{\Psi}(z)=\sum_n\op{a}_{n}  \psi_{n}(z);\h{1cm}  \psi_{n}(z)=\frac{1}{\sqrt{2\pi R}}e^{in(z/R)},\en
the Hamiltonian becomes
\begin{eqnarray}
\op{H}&=& \sum_{n} \hbar\omega_n\op{a}_{n}^{\dag}\op{a}_{n} + \frac{1}{2}\hbar\chi\sum_{n+k-l-s=0} \op{a}_{n}^{\dag}\op{a}_{k}^{\dag} \op{a}_{l} \op{a}_{s}
 \nonumber \\
& & +\hbar\left[u_-\op{a}_{n}^{\dag}\op{a}_{n-2q} e^{-i2q\omega t}+u_+\op{a}_{n}^{\dag}\op{a}_{n+2q} e^{i2q\omega t} \right],
\label{Hama}
\end{eqnarray}
where we have defined the effective 1D interaction strength $\chi=\frac{g}{4\hbar\pi^2 l_r^2R} $,
unperturbed
eigenenergies $\hbar\omega_n=\frac{\hbar^2 n^2}{2mR^2}$ and potential amplitudes $u_\pm=\frac{1}{2}(u_x\pm iu_y)$.  The equations of motion for the operators $a_n$ are
\begin{eqnarray}
i\frac{\partial}{\partial t}\op{a}_{n}(t)
&=&(\omega_n-n\omega)\op{a}_{n}+\left[u_-\op{a}_{n-2q}+ u_+\op{a}_{n+2q}\right]\n\\&&
+\chi\sum_{k}\sum_{l}\op{a}^\dagger_{k}
\op{a}_{l} \op{a}_{n+k-l}.
\label{rotating-Schr-equation}
\end{eqnarray}
Here we have redefined the operators by replacing $\op{a}_{n}(t)\rightarrow \op{a}_{n}(t)e^{-in\omega t}$ to remove the explicit dependence on time in the equations, which is equivalent to transforming  to co-ordinates co-rotating with the lattice potential with its angular frequency $\omega$.

If the energy gap $\hbar\omega_n$ is large then the coupling among different energy states can be neglected and the dynamics can be restricted to the subspace of the two degenerate modes $\pm q$ that match lattice periodicity. In that scenario, we denote $ \op{a}_{q} =\op{a}$ and $ \op{a}_{-q} =\op{b} $, governed by two coupled dynamical equations
\begin{eqnarray}
 i \dot{\op{a}} &=& -q\omega \op{a} + u_-\op{b}
+  \chi \left(\op{a}^{\dag} \op{a}^{2} + 2\op{a} \op{b}^{\dag}\op{b}\right) ,
\n\\
 i \dot{\op{b}} &=&  q\omega \op{b} + u_+\op{a}
+  \chi \left(2\op{a}^{\dag} \op{a} \op{b} +  \op{b}^{\dag}\op{b}^2 \right)
 .
\label{two-level-rotating-Schr-equation-nonlin}
\end{eqnarray}
The corresponding effective 2-mode Hamiltonian
\begin{eqnarray}
\op{H}_{2m}&=&- \hbar q \omega \left(\op{a}^{\dag} \op{a} -  \op{b}^{\dag} \op{b}\right)+
\hbar u_-\op{a}^{\dag} \op{b} +\hbar u_{+}\op{a}  \op{b}^{\dag}
\nonumber  \\
& & + \frac{\hbar \chi}{2} \left(\op{a}^{\dag 2} \op{a}^{2} + 4
\op{a}^{\dag} \op{a}  \op{b}^{\dag} \op{b} +  \op{b}^{\dag 2} \op{b}^{2} \right) ,
\label{Ham-ab}
\end{eqnarray}
is time-independent.

For some of our analysis, we will find it useful to express this two-mode Hamiltonian in terms of pseudo-spin operators
\begin{eqnarray}
\op{J}_x &=& \frac{1}{2}(\op{a}^{\dag}\op{b}+\op{a}\op{b}^{\dag}), \h{1cm}
\op{J}_y = \frac{1}{2i}(\op{a}^{\dag}\op{b}-\op{a}\op{b}^{\dag}), \n\\
\op{J}_z &=& \frac{1}{2}(\op{a}^{\dag}\op{a}-\op{b}^{\dag}\op{b}), \h{1cm}
\frac{N}{2} = \frac{1}{2}(\op{a}^{\dag}\op{a}+\op{b}^{\dag}\op{b}),
\label{eqabN}
\end{eqnarray}
satisfying the commutation relations
$[\op{J}_i,\op{J}_j]=i\epsilon_{ijk}\op{J}_k$, where $\epsilon_{ijk}$ is the Levi-Civita symbol and Einstein summation convention is assumed,  and
\begin{eqnarray}
\op{J}_x^2 + \op{J}_y^2 +  \op{J}_z^2  &=&
\frac{N}{2} \left(\frac{N}{2} + 1\right).
\label{eqabN2}
\end{eqnarray}
The Hamiltonian can then be written as
\begin{eqnarray}
\hat{H}_{2m} &=& \hbar \left(u_x \op{J}_x + u_y \op{J}_y  - 2q\omega \op{J}_z  - \chi \op{J}_z^2 \right) ,
\label{HamJ}
\end{eqnarray}
where we have dropped a c-number term $\chi \left(\frac{3}{4}N^2 -\frac{1}{2}N \right)$ that produces a global phase, not relevant to the two-mode dynamics.

For our numerical simulations, we will use energy, angular frequency and time units of $\epsilon=\hbar\omega_0=\hbar^2/(mR^2)$ and $\tau=2\pi/\omega_0$,and use $q=5$. In the rest of the paper, we will generally set $u_y=0$ and use single lattice strength parameter $u$ defined by  $u_+=u_-=u_x/2=u$.

\section{Mean field dynamics}

Much of the essential dynamics of this model can be understood within a mean field approximation, where quantum correlations and fluctuations
are neglected. This amounts to replacing the operators by their expectations $\op{a}_n\rightarrow\langle\op{a}_n\rangle=a_n$, $\op{J}_k\rightarrow\langle\op{J}_k\rangle=J_k$.
This approximation is useful when dealing with large total particle number $N$ and condensates that are weakly interacting. One can make the analogy of this approximation with transition from the quantum electrodynamics to classical electromagnetism. Within the mean field picture, we examine separately the dynamics in the absence and presence of inter-atomic interactions.

\subsection{Non-interacting linear limit}

\begin{figure}[t]
\centering
\includegraphics[width=\columnwidth]{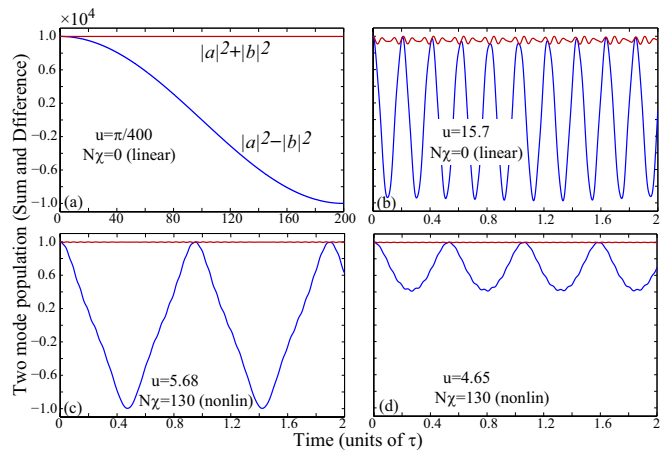}
\caption{(Color online) Time evolution of the sum and difference of  populations $|a(t)| ^2$ and $|b(t)| ^2$ in states $\ket{\pm q}$, is plotted by numerically solving Eq.~(\ref{Psi-evolution}). The constancy of their sum justifies the two-mode model Eqs.~\eqref{amotion4} when $u\ll \hbar^2q^2/2mR^2$. Plots (a) and (b) illustrate the non-interacting case; (b) shows that significant fluctuations out of two modes occurs  only at much higher lattice potentials ($u$ is 2000 times higher in (b) compared to (a)). With nonlinearity, there can be both (c) oscillatory and (d) self-trapped regimes.}
\label{Figure-2}
\end{figure}

It is instructive to first consider the non-interacting limit, setting $\chi=0$, so that in the two mode approximation the equations reduce to
\begin{eqnarray}
 i \dot{a} &=& -q\omega a + u b \h{1cm}
 i \dot{b} = q\omega b + u a.
\label{amotion4}
\end{eqnarray}
These equations have exact analytical solutions,
\bn
\label{solution-two-level-rotating-Schr-equation}
a( t)=[ \cos(\eta t)+i\frac{\omega q}{\eta}\sin(\eta t)]a(0) -i\frac{u}{\eta}\sin(\eta t)b(0), \hspace{.3cm}\\
 b( t)= -i\frac{u}{\eta}\sin(\eta t) a(0)+  [ \cos(\eta t)-i\frac{\omega q}{\eta}\sin(\eta t)]b(0),\n
\en
where we defined $\eta=\sqrt{u^2+\omega^2q^2}$. The lattice potential $u$ couples the two modes, while the angular velocity $\omega$ lifts their degeneracy. We choose our initial state to be $|a(0)|^2=N$, $|b(0)|^2=0$. With no rotation ($\omega=0$) Eq.~\eqref{solution-two-level-rotating-Schr-equation} implies a complete population swap between the states at time $t_s=\pi/2u$, i.e.  $|a(t_s,\omega=0)|^2=0$, $|b(t_s,\omega=0)|^2=N$ (see  Fig.~\ref{Figure-2}a). For $\omega\neq 0$ the initial state remains partially populated,
\begin{eqnarray}
|a( t_s)|^2&=&N\left[\cos^2\left(\frac{\pi}{2}\frac{\eta}{u} \right)
+\frac{\omega^2q^2}{\eta^2} \sin^2\left(\frac{\pi}{2}\frac{\eta}{u} \right)\right],
\label{probability}
\end{eqnarray}
due to the lifted degeneracy of the two states in the rotating frame.  The
dependence of $|a(t_s)|^2$ on the rate of rotation $\omega$ is plotted in Fig.~\ref{Figure-3}(a) for the same parameters as in Fig.~\ref{Figure-2}(a). It suggests that the mode population after a fixed evolution time can be used for rotation frequency measurements.


\subsection{Nonlinear regime}
To examine the effects of the non-linear terms arising from interactions, we first perform  numerical simulations \emph{without} making the two-mode approximation. We  solve the coupled mean field equations
\begin{eqnarray}
i\frac{\partial}{\partial t}a_{n}(t)
&=&(\omega_n-n\omega)a_{n}+\left[u_-a_{n-2q}+ u_+a_{n+2q}\right]a_{n}\n\\&&
+\chi\sum_{k}\sum_{l}a^*_{k}
a_{l} a_{n+k-l}.
\label{Psi-evolution}
\end{eqnarray}
using a range of values of $n$ about the resonant modes $\pm q$. The results shown in Fig.~\ref{Figure-2} confirm that for both the linear and nonlinear regimes, the two-mode approximation is quite accurate provided that the lattice strength remains small, $u\ll \hbar^2q^2/8mR^2$. When that condition is satisfied, the population remains in the two-mode subspace as indicated by the constancy of the sum $|a(t)|^2+|b(t)|^2$ during time evolution  (Fig.~\ref{Figure-2}a, c, and d). Deviations occur only when $u$ is large as in Fig.~\ref{Figure-2}b.  Therefore, for most cases the two-mode approximation can be applied.

The nonlinearity leads to an additional time-dependent potential due to the atomic density pattern.  In the two-mode picture, this is proportional to  $[ab^*\exp(i2q({\textstyle\frac{z}{R}}-\omega t))+ c.c.]$, and its effects can be clearly seen if the two-mode equations are written in the form
\begin{eqnarray}
 i \dot{a} &=& -q\omega a + u b
-\chi |a|^2 a  ,
\n
\\
 i \dot{b} &=&  q\omega b + u a
-\chi |b|^2 b ,
\label{amotion4}
\end{eqnarray}
where $a$ and $b$ have been redefined to absorb a global phase. This shows that the nonlinear terms appear diagonally and add/subtract to the rotation term, thereby influences the sensitivity to the rotation.

%
\begin{figure}[t]
\centering
\includegraphics[width=\columnwidth]{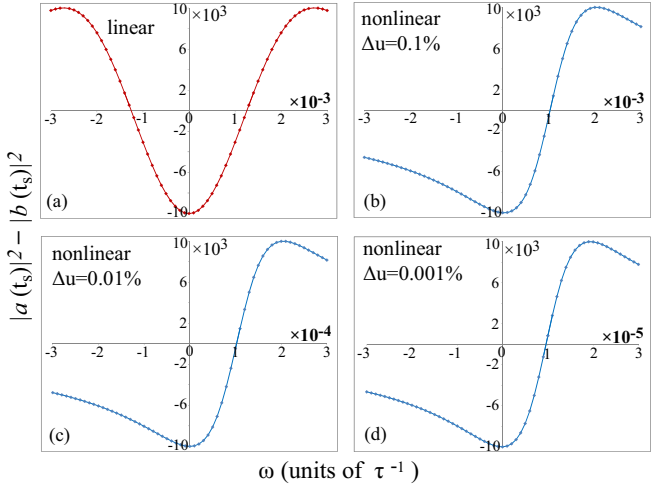}
\caption{(Color online) In the mean-field approximation, starting with $|a(0)|^2=N$, the time-evolved population $|a( t_s)|^2$ in mode $|+q\rangle$ at the fixed time $t_s$ is plotted as a function of the angular velocity $\omega$; here $t_s$ is the time required for a complete swap to state $|-q\rangle$, $|b( t_s)|^2=N$ in the non-interacting case when $\omega=0$ as seen in (a). The remaining panels (b-d) show that nonlinearity sharpens the dependence on the $\omega$, as $u\rightarrow u_c=N\chi/4$, the critical value from above with $\Delta u=100(u-u_c)/u_c$. Although the plots look similar,
note the scale of the horizontal axis that indicates two orders of magnitude increase in sensitivity to changes in $\omega$ as criticality is approached.}
\label{Figure-3}
\end{figure}

Insight into the dynamics of the system can be gained from a semiclassical picture where the operators are replaced with c-number variables parameterized as
\begin{eqnarray}
J_x &=& \frac{N}{2} \sqrt{1-z^2} \cos \phi, \n\\
J_y &=& \frac{N}{2} \sqrt{1-z^2} \sin \phi, \n\\
J_z &=& \frac{N}{2} z ,
\end{eqnarray}
where the dimensionless dynamical variables have the range $-1 \leq z \leq 1$ and $-\infty \leq \phi \leq \infty$. Thereby, the Hamiltonian in Eq.~(\ref{HamJ}) reduces to the semi-classical form
\begin{eqnarray}
H_{sc}=-\frac{N\hbar}{2}\left(\frac{N\chi}{2}z^2-2u\sqrt{1-z^2}\cos\phi+2q\omega z\right).
\label{Hsc}
\end{eqnarray}
The canonically conjugate pair of variables $\phi$ and $\frac{N\hbar z}{2}$ yield the equations of motion
\begin{eqnarray}
\dot{z} &=& -\frac{2}{N\hbar} \frac{\partial H}{\partial \phi} \nonumber \\
 &=& 2u \sqrt{1-z^2} \sin \phi, \\
\dot{\phi} &=& \frac{2}{N\hbar} \frac{\partial H}{\partial z} \nonumber \\
&=& -\frac{2uz}{\sqrt{1-z^2}}\cos \phi -2q\omega - N\chi z.
\end{eqnarray}
The Hamiltonian (\ref{Hsc}) corresponds to a ``nonrigid pendulum'' dynamics introduced as a model of quantum coherent atomic tunneling between two trapped BECs \cite{Smerzi-semiclassical}.

\begin{figure}[t]
\centering
\includegraphics[width=\columnwidth]{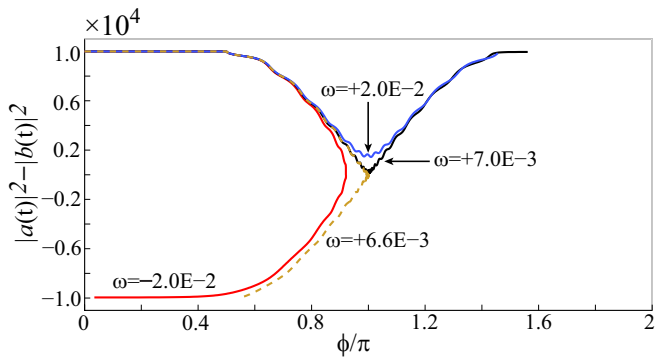}
\caption{(Color online)  Phase space trajectories of the semiclassical dynamical variables $z$ and $\phi$.
Each line represents time evolution for specific values of $\chi$ and $\omega$. Transition  from oscillatory to self-trapped behavior as $\omega$ is varied with the interaction strength $\chi$  \emph{fixed} (just below the critical point for $\omega=0$).}\vspace{-.5cm}
\label{Figure-4}
\end{figure}

It is convenient to introduce a parameter $\Lambda = N\chi/(2u)$ characterizing the ratio between the nonlinear interaction and the Rabi oscillations, as in \cite{Zibold2010}. The Hamiltonian then takes the form
\begin{eqnarray}
H_{sc}=-\frac{N\hbar u}{2}\left(\Lambda z^2-2\sqrt{1-z^2}\cos\phi+\frac{2q\omega z}{u}\right),
\label{Hsc2}
\end{eqnarray}
which can be expanded in the vicinity of  $z=0,\phi=\pi$ up to the second order in $dz,d\phi$ as
\begin{eqnarray}
H_{sc}\approx -\frac{N\hbar u}{2}\left[2+ (\Lambda-1) dz^2+\frac{2q\omega }{u} dz -d \phi^2\right].
\label{Hsc3}
\end{eqnarray}
Changing the nonlinearity parameter from $\Lambda <1$ to $\Lambda >1$ the trajectories near  $z=0,\phi=\pi$ change from elliptical to hyperbolic. Being in the hyperbolic regime,  $\Lambda >1$, we are interested in the trajectory that goes through the pole, $z=1$. This corresponds to the value of the Hamiltonian $H_{sc} = -\frac{N\hbar u}{2}(\Lambda +2q\omega/{u})$, yielding
\begin{eqnarray}
\frac{\Lambda - 2 + \frac{2q\omega}{u}}{\Lambda - 1} = \left[ dz+\frac{q\omega}{u(\Lambda - 1)} \right]^2
-\frac{d\phi ^2}{\Lambda - 1} .
\label{Hsc4}
\end{eqnarray}
Sign of the expression on the left hand side of Eq. (\ref{Hsc4}) determines whether the hyperbolic trajectory crosses to the opposite hemisphere or becomes self-trapped. For this change, the critical value of the nonlinearity corresponds to
\begin{eqnarray}
\Lambda_{\rm crit} = 2-\frac{2q\omega}{u} .
\end{eqnarray}
Choosing the Rabi frequency $u=u_c\equiv N\chi/4$, we see that the transition happens for frequency $\omega = 0$ where $\Lambda_{\rm crit} = 2$. The corresponding change of behavior can be seen in Figs. \ref{Figure-2}(c) and (d),  \ref{Figure-3} and  \ref{Figure-4}.

Although the equations of motion are identical in form to those describing BECs in coupled double wells \cite{Smerzi-semiclassical,wang}, the interpretation is different:
(i) The relevant states are circulating modes that occupy the same physical space, (ii) the lattice provides the coupling, (iii) the role of the onsite energies is played by the rotation,
(iv) the nonlinear term has negative $-\chi \op{J}_z^2$ in Eq. (\ref{HamJ}) despite a repulsive interaction. This last point, seemingly contradictory, can be explained \cite{OKD2015} as follows: For $J_z \approx 0$ the counter-rotating modes $a$ and $b$ are almost equally populated and the condensate forms a standing wave with pronounced interference fringes. Therefore, the particles are effectively compressed to half of the volume that would be occupied if they were all circulating in the same direction $(J_z \approx \pm N/2)$ with no interference fringes. Thus, states with larger $|J_z|$ correspond to lower interaction energy .

An essential distinguishing feature of our specific system emerges when $\omega\neq 0$: Then even when the nonlinearity and the lattice strength are constant, simply varying the rate of rotation can lead to a transition from self-trapping to oscillatory behavior as shown in Fig.~\ref{Figure-4}. Thus, when the system is close to criticality the nonlinearity can significantly enhance sensitivity to the rotation by magnifying the dependence of the observable $|a(t_s)|^2$ on $\omega$. Specifically, for a fixed potential $u$, and for $\omega=0$, consider that the nonlinearity is tuned to be just below the critical point in the oscillatory regime and the time required for complete swap $t_s$ is set to be the detection time. When rotating the system ($\omega\neq 0$) there are two possibilities depending on the direction of rotation: (i) In one direction, the rotation tips the system over into the self trapped-regime; (ii) in the opposite direction the system stays in the oscillatory regime, but due to the proximity to the critical point there is still a high sensitivity to small changes in $\omega$.  Note that changing the sign of the nonlinearity switches the dependence on the direction of rotation in the relevant regime close to $\omega=0$, as follows from the relative signs of $\omega$ and $\chi$ in  Eqs.~\eqref{amotion4}.

Figure~\ref{Figure-3} illustrates this enhancement of the sensitivity to rotation due to nonlinearity by about two orders of magnitude compared to the linear case, as the lattice strength approaches the critical value $u\rightarrow u_c=N\chi/4$ for fixed nonlinearity $\chi$. An additional feature implicit in the parameters shown in the plots is that the associated values of the lattice potential $u$ can be substantially higher than in the linear case, which has the effect of reducing the swap time $t_s$ by about two orders of magnitude. Thus, nonlinearity not only increases the sensitivity, but the process can be accomplished in shorter times as well. Nevertheless, to get useful metrological applications, one also has to take into account the noise generated in the process.
One source of noise stems from fluctuations of the total atomic number $N$: if $u$ is chosen assuming the atomic number to be $N_0$, but the actual number is $N=N_0+\Delta N$, the critical value of the frequency is changed to $\omega_c=-\chi \Delta N/(4q)$. This limits the precision with which $\omega$ can be measured. Another source of noise are quantum fluctuations which are enhanced especially in the vicinity of critical values of the parameters. The corresponding dynamics is studied in the next section.


\section{Quantum dynamics}

\subsection{Full quantum simulations}

The mean-field simulation demonstrates the existence of critical behavior, however, near criticality quantum fluctuations also get enhanced. Since mean field does not account for that, we now examine the quantum dynamics of the system. Based on our results in the previous section we restrict our analysis to a two mode subspace.
The Hilbert space is spanned by the Fock states $\ket{n_a,n_b}\equiv \ket{N-n,n}$, hence the general state of the atoms is
\begin{eqnarray}
\ket{\Phi(t)}=\sum_{n=0}^N c_n(t)\ket{N-n,n}.
\label{discreet-state}
\end{eqnarray}
For this state, we can quantify the fluctuations of the operator $\op{J}_i$, $i=x,y,z$, by the standard deviation
\begin{eqnarray}
\sigma_i\equiv \sqrt{\langle \op{J}_i^2\rangle-\langle \op{J}_i\rangle^2},\;i=x,y,z.
\label{discreet-deviation}
\end{eqnarray}

\begin{figure}[t]
\centering
\includegraphics[width=\columnwidth]{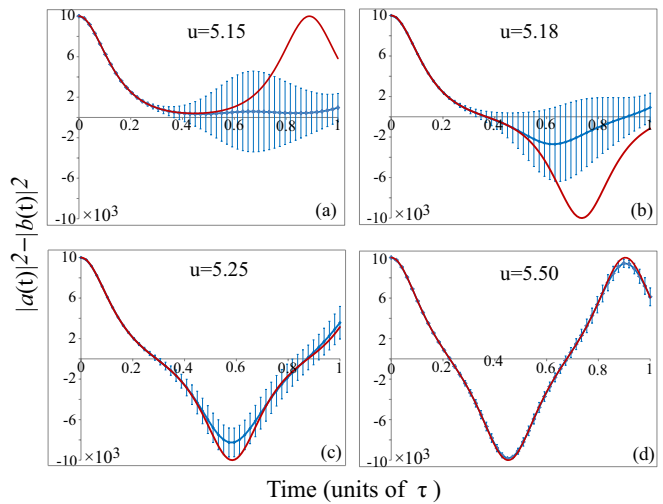}
\caption{(Color online)  Comparison of the two-mode quantum time-evolution (blue line with error bars) with the mean field evolution (red lines without error bars). The population difference between the two modes of interest are plotted as a function of time for no rotation $\omega=0$. The transition from self trapping to oscillatory behavior is seen as the lattice strength
$u$ passes through the critical value of $u_c=N\chi/4=5.155$. Notably close to the critical point, the quantum fluctuations are enhanced and there is substantial difference between the two pictures, but not so further from the critical point.}
\label{Figure-5}
\end{figure}

We do our simulations for $N=10^4$ atoms for quantum evolution, with $u_x=2u$ and $u_y=0$,
and compare the results with the mean field dynamics in Fig.~\ref{Figure-5}. We plot the population difference between the two states, which in the quantum simulation is $\langle 2\hat{J}_z\rangle$. We also indicate the quantum fluctuations by vertical bars measured by $\pm \sigma_z$ (this is consistent with a minimum resolution of $2\sigma_z$ utilized in Sec.~VI). There is qualitative agreement between the mean-field approximation and the quantum dynamics. Specifically, the panels in
Fig.~\ref{Figure-5}
show similar critical behavior in both approaches, when the lattice depth $u$ is varied keeping the interaction strength $\chi$ fixed: There is transition from self-trapped
Fig.~\ref{Figure-5}(a)
to oscillatory
Fig.~\ref{Figure-5}(b,c,d) regime
which even for the quantum dynamics occurs in the vicinity of the critical ratio $N\chi=4u$ predicted by the
mean field theory. Away from the critical value $u_c=N\chi/4$, the quantum dynamics is almost identical to the mean field dynamics, remarkably even for relatively strong non-linearity assumed here, showing that only weak inter-atomic correlations are created during the evolution.

However, pronounced differences emerge close to the criticality. The transition from self-trapping to oscillatory is more gradual in the quantum dynamics as can be seen by the progression from Figs.~\ref{Figure-5}(a) through (d).  One can interpret the differences as a smearing effect brought on by the quantum fluctuations which are small far from the critical point but get progressively larger as the critical point is approached.  In the quantum dynamics the time and magnitude of the maximum swap also differ from the mean field results, with the maximum swap being incomplete in the quantum case.

\begin{figure}[t]
\centering
\includegraphics[width=\columnwidth]{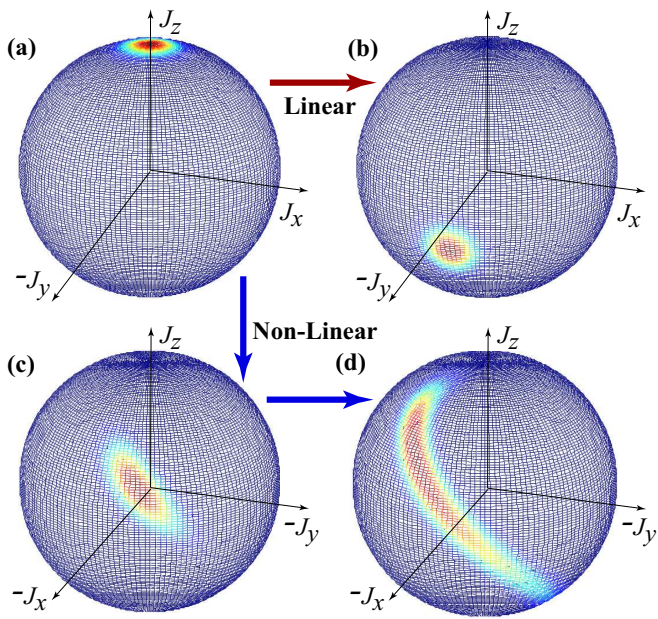}
\caption{(Color online)  Bloch sphere representation of the two-mode quantum mechanical time-evolution with (a) initially all the population centered at the north pole ($\langle|a(0)|^2\rangle=N$). (b) In the absence of interaction, the density distribution undergoes rotation but its shape is maintained. (c) With interaction, in additon to rotation, spin-squeezing distorts the distribution, and (d) which eventually destroys its Gaussian profile as well. Here we use $N=100$ and $N\chi=2.06$ for illustration, but for $N=10,000$ used in our estimates squeezing is much more pronounced. }
\label{Figure-6}
\end{figure}

The dynamics is best visualized on Bloch sphere as shown in Fig.~\ref{Figure-6} where
$Q$-functions of the states are plotted. The $Q$-function is the squared projection of the state on the spin coherent state with the mean values of operators $\hat{J}_{x,y,z}$ defining position on the Bloch sphere.
The initial state is a spin coherent state with $\langle \hat{J}_{z} \rangle = N/2$ with a $Q$-function of a Gaussian-like blob centered at the north pole.  In the linear case ($\chi=0$) the distribution maintains its shape which for $\omega=0$ rotates around the $\hat{J}_x$ axis so that south pole is reached at $ut=\pi/2$.

The nonlinear term introduces twisting around the $J_z$ axis.
Combination of rotation around $J_x$ and twisting around $J_z$ leads to a more complicated migration of the blob. For critical values of the parameters at the center of the blob approaches the  $-J_x$ axis where it stays, whereas for even stronger nonlinearity the blob gets to a trajectory returning back to the north pole.
The twisting also deforms the blob as it is being squeezed in one direction and stretched in the other: This is the spin squeezing effect first proposed by Kitagawa and Ueda  \cite{Kitagawa}. As a result, for some variables we can find suppressed noise whereas for others  the noise is amplified. Specifically, in Fig.~\ref{Figure-6} the orientation of the stretching indicates amplified fluctuations in $J_z$, hence in the modal population difference.

\begin{figure}[t]
\centering
\includegraphics[width=\columnwidth]{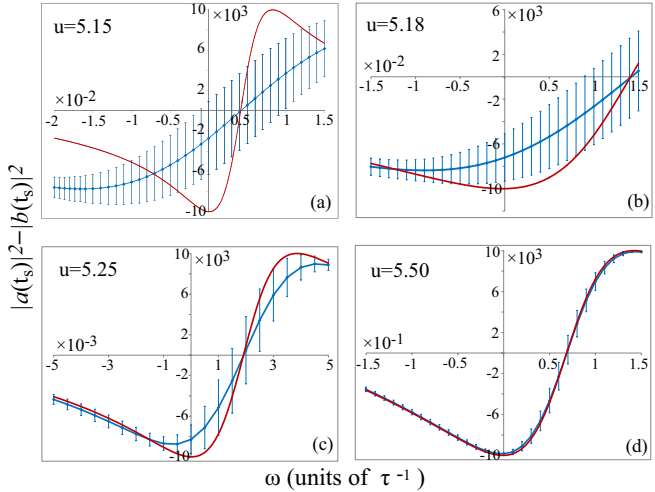}
\caption{(Color online)  Comparison of the two-mode quantum time-evolution ((blue line with error bars) with the mean field evolution (red line without error bars)). The population difference between the two modes of interest as a function of the rotation rate $\omega$ after fixed time evolution $t_s$, the time for full swap with no interaction and no rotation. Notably close to the critical point $u_c=N\chi/4=5.155$, the quantum fluctuations are enhanced and there is substantial difference between the two pictures, but not so further from the critical point.}
\label{Figure-7}
\end{figure}

This effect directly impacts the sensitivity to rotation close to criticality. We illustrate this in Fig.~\eqref{Figure-7}, where the plots are analogous to those in Fig.~\eqref{Figure-3}, but now along with the mean-field evolution, the results of our quantum mechanical simulation are plotted including fluctuations. The fluctuations are large near criticality [panels (a) and (b)] whereas far from criticality the mean field approximation is almost indistinguishable from the quantum simulation.  Near criticality there is also a significant difference in the location of the minimum (the point of maximum swap). These effects can be explained as results of bifurcating phase trajectories near the intersection of the Bloch sphere with the $J_x$ axis.

\subsection{Gaussian Approximation of Quantum dynamics}

Numerical modeling of the quantum mechanical evolution becomes quite demanding for large $N$, therefore, approximate solutions are helpful. For short times one can assume that the state remains approximately Gaussian and find a closed set of nine equations for the first and second moments ${\cal J}_i$ and $V_{ij}$ of $\hat{J}_{i}, (i=x,y,z)$ given in  Appendix A \cite{Vardi2001,TO-tensor}.
We solve these equations with $u_x=2u$ and $u_y=0$ and with the initial conditions
\begin{eqnarray}
&&{\cal J}_{z,0} = \frac{N}{2}, \quad  V_{xx,0} =  V_{yy,0} =  \frac{N}{4}, \\
&&{\cal J}_{x,0} = {\cal J}_{y,0} = V_{zz,0} =  V_{yz,0} = V_{xy,0} =  V_{xz,0} =0,\n
\end{eqnarray}
i.e., starting from a spin-coherent state located at the north-pole of the Bloch sphere.
\begin{figure}[t]
\centering
\includegraphics[width=\columnwidth]{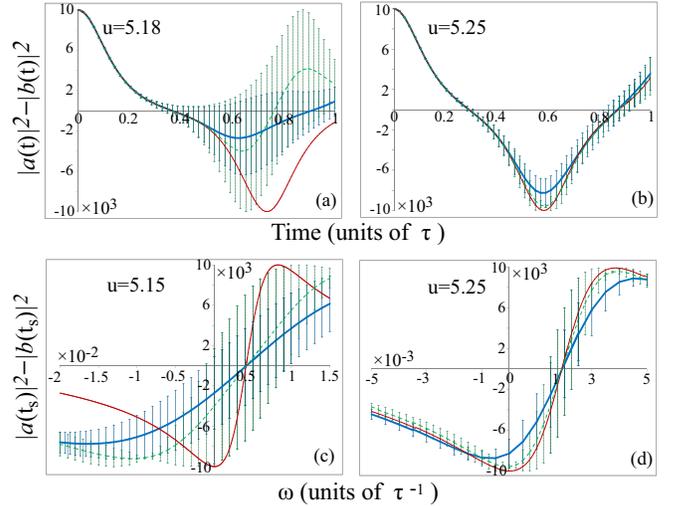}
\caption{(Color online)  Comparison of the Gaussian approximation (green dashed line with dotted error bars) with quantum simulation (thick blue line with solid error bars) as well as mean-field approximation (thin red line without error bars). The upper two plots show the time evolution of the modal population difference, corresponding to Fig.~\ref{Figure-5}(b) and (c), and the lower two plots its value at time $t_s$ as  function of the rotation rate $\omega$, and corresponds to Fig.~\ref{Figure-7}(a) and (c). The Gaussian approximation generally overestimates the fluctuations more so closer to criticality. }
\label{Figure-8}
\end{figure}

The comparison between the evolution of moments calculated from the approximate equations Eqs.~(\ref{eqJx})--(\ref{eqVyz}) is in Fig.~\ref{Figure-8}  with the upper panels displaying the time evolution and the lower panels the impact of rotation.  The results of the Gaussian approximation including fluctuations are overlaid with their equivalents calculated from the two mode quantum analysis and mean field approximation. There is a good agreement among all three methods for short times and parameters far  from the critical point [Figs.~\ref{Figure-8} (b) and \ref{Figure-8}(d)].  Even when the parameters are chosen close to criticality, the time evolution [panel (a)] in the Gaussian approximation matches the quantum simulation quite well in the approach to the equator in the Bloch sphere since the distribution maintains the Gaussian shape. Beyond that point, for the critical parameters, there is significant difference because the $J_z$ twisting distorts the distribution [see Fig.~\ref{Figure-6} (d)] from a Gaussian shape, so the Gaussian approximation tends to overestimate the fluctuations beyond the approach to the critical point.  The Gaussian approximation however serves the purpose of confirming the two-mode quantum dynamics as can be seen in their general similarity in the $\omega$-dependence in panels Figs.~\ref{Figure-8}(c) and \ref{Figure-8}(d).


\section{Squeezed Ramsey Interferometry}

\begin{figure}[t]
\includegraphics[width=\columnwidth]{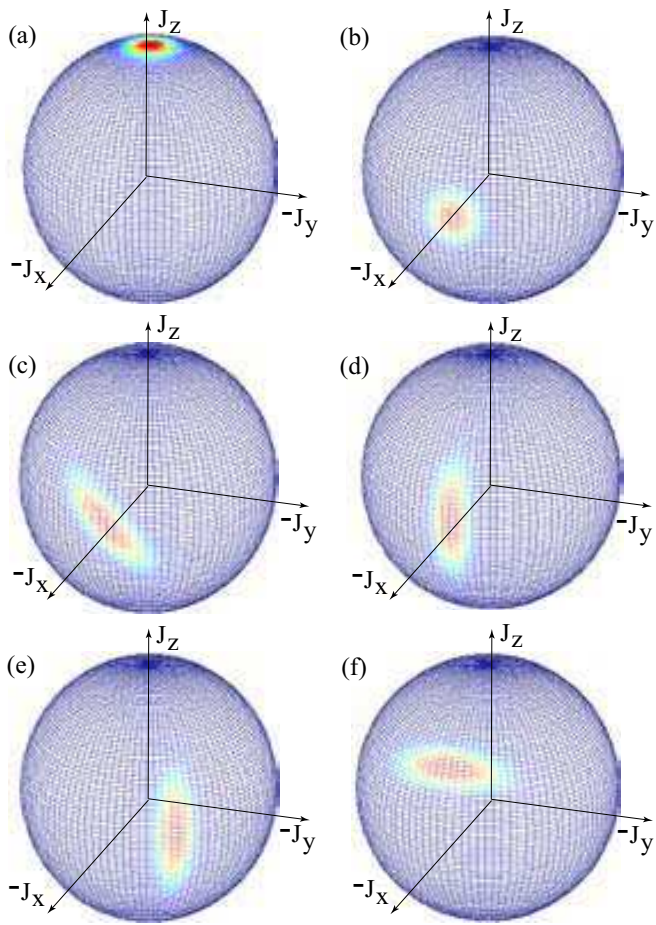}
\caption{(Color online) The sequence of steps in the Ramsey interferometric scheme to suppress fluctuations: (a) Initial state with mean population entirely in state $a=|+q\rangle$, (b) the state is transformed to be centered on the equator, (c) the state has undergone nonlinear evolution with maximal spin squeezing, (d) the state is transformed to have long axis parallel to the $\hat{J}_z$ axis, (e) free evolution solely under influence of angular velocity $\omega$, (f) the state is transformed to have short axis aligned with the quadrature of interest $\hat{J}_z$ for reduced fluctuations. Here we use $N=100$  and $N\chi=10.3$ for illustration; but for $N=10,000$ as used in our numerical estimates squeezing is much more pronounced.}
\label{Figure-9}
\end{figure}

The quantum simulations have shown that fluctuations due to squeezing near criticality effectively erase
the enhanced rotation sensitivity indicated by a mean field picture. However, the problem carries within it a solution as well, because the squeezing can also be used to suppress fluctuations in interferometric schemes to improve measurement precision \cite{Wineland1994,Lloyd1,Lloyd2}. Here we describe a Ramsey-style interferometric sequence \cite{Wineland-PRA1992} applied to the present orbital model as  steps labeled (a-f)  in  Fig.~\ref{Figure-9}.

(a -b) The initial state is prepared as the $J_z=N/2$ coherent state on the pole of the Bloch sphere (all atoms orbiting in one direction). Application of the $J_y$ operator  rotates the state to the equator centered on  $-J_x$ corresponding to a superposition of atoms orbiting in each of the two directions. The superpositions is balanced with approximately equal atomic numbers in the two directions, with  binomial fluctuations $\sqrt{N}/2$. The density distribution forms an interference pattern along the ring with fringes positioned with angular precision  $\sim 1/q\sqrt{N}$.

(b-c) The nonlinear Hamiltonian is applied
\begin{eqnarray}
\op{H}_c=-2\omega q \op{J}_z+2u \op{J}_x-\chi\op{J}_z^2,\ \ u=\frac{N\chi}{4}
\label{discreet-hamiltonian-squeezing}
\end{eqnarray}
for duration $ t_{\rm squeeze}$ to generate squeezing, for short times described by the squeezing parameter $\xi^2 \approx\exp(-\chi N t_{\rm squeeze})$ ($\xi^2$ is the ratio of the minimum variance on the Bloch sphere to the variance of the spin coherent state) \cite{TO-tensor,Wineland-PRA1992}. The lattice strength $ u=N\chi/4$ is chosen to be at the optimum value ensuring that the Bloch-sphere distribution is kept aligned  along the $-\pi/4$ orientation centered at the $-\hat{J}_x$ axis for fastest squeezing \cite{TO-tensor}.
Note that although this value corresponds to the critical value of $\Lambda=2$ for which the classical trajectories going through a pole change from oscillatory to self-trapped, here the precision of $u$ is not critical. Choice of $u=N\chi/4$ leads to fastest squeezing generation, but deviations from the optimum value caused, e.g., by fluctuations of $N$, do not influence substantially precision of measurement of $\omega$.
The duration of $ t_{\rm squeeze}$ is chosen to achieve maximum squeezing value; longer duration would degrade the squeezing due to distortion of the distribution into an $S$-shaped form, displayed in Fig.~\ref{Figure-6} (d). During the squeezing process fluctuations of both the atomic number difference and of the fringe position increase, however their correlations become stronger.

(c-d) When the maximally squeezed state is reached, the nonlinearity $\chi$ is switched off by suppressing the atomic interaction, possibly by Feshbach resonance, changing the Hamiltonian to a linear form
\begin{eqnarray}
\op{H}=-2\omega q \op{J}_z+2u \op{J}_x.
\label{discreet-hamiltonian-transformation}
\end{eqnarray}
This Hamiltonian acts for $ t_{\rm transform}=\pi/(8 u)$ to rotate the state about the $-\hat{J}_x$ axis until the {\em long} axis of the distribution is aligned along $J_z$ direction. In this state the fluctuations of the atomic number difference become large, however the fluctuations of the position of the fringes become suppressed below the coherent state value $\sim 1/q\sqrt{N}$.

(d-e) The lattice potential is switched off for $t_{\rm sensing}$ so that the Hamiltonian is
\begin{eqnarray}
\op{H}_c=-2\omega q \op{J}_z,
\label{discreet-hamiltonian-sensing}
\end{eqnarray}
and the state evolves purely under the rotation of the frame, thus acquiring the interferometric phase.
The sharply localized fringes now change their position by $2q\omega t_{\rm sensing}$.

(e-f) The lattice is restored to the same strength as in step (c-d)
\begin{eqnarray}
\op{H}_c=-2\omega q \op{J}_z+2u \op{J}_x,
\label{discreet-hamiltonian-transformation2}
\end{eqnarray}
and the state evolves under its influence for $ t_{\pi/2}=\pi/(4u)$ equivalent to a  $\pi/2$ pulse  to rotate the state to be now with {\em short} axis along $\hat{J}_z$ direction. Then the number of atoms in both modes is measured to find $J_z$. The atomic number difference now has suppressed fluctuations and contains information on the rotational frequency $\omega$.

Results of our simulation of this sequence are summarized in Table~I for two different particle numbers $N=1,000$ and $N=10,000$. We find that rotation of the squeezing axis reduces fluctuation $\sigma_z$ from $81.6$ to $3.24$ and from $480$ to $5.65$ for the two cases respectively, and in both cases, they represent significant suppression of fluctuations over the binomial values of $15.8$ and $50.0$ in the non-interacting linear case.

\begin{table}
\caption{The evolution of the fluctuations in the population difference as measured by the standard deviation in $\langle J_z\rangle$ is tabulated for the various steps of the Ramsey interferometric scheme shown in Fig.~\ref{Figure-9}. Results of simulations for two different values of total particle number $N$ are displayed. The final reduced fluctuations should be compared with the fluctuations for the non-interacting linear case, both shown in bold letters. The beginning of step (b-c) is taken as $t=0$.}
\begin{tabular}{ccccccccc}
 &\vline & N &=   &   1,000 &\h{1cm}\vline \vline  & N &=  & 10,000\\
\hline \hline
step &\vline &t ($\tau$)&\vline   &   $\sigma_z$ &\h{1cm}\vline \vline  & t ($\tau$) &\vline   & $\sigma_z$\\
\hline\hline b &\vline & 0 &\vline   &    15.8 &\h{1cm}\vline  \vline & 0 &\vline   & 50.0\\
\hline c &\vline & 1.61 &\vline   &   58.3&\h{1cm}\vline  \vline & 0.220 &\vline   & 340\\
\hline d &\vline &  2.37 &\vline   &   81.6 &\h{1cm}\vline  \vline & 0.296 &\vline   & 480\\
\hline e &\vline &  3.37 &\vline   &   81.6 &\h{1cm}\vline  \vline & 1.296 &\vline   & 480\\
\hline f &\vline & 4.89 &\vline   &   {\bf 3.24} &\h{1cm}\vline  \vline & 1.446 &\vline   & {\bf 5.65}\\
\hline\hline  &\vline & linear &\vline   &  {\bf 15.8} &\h{1cm}\vline  \vline & linear &\vline   & {\bf 50.0}
\\
\hline \\
\end{tabular}
\end{table}

We conclude this section by noting, that the Ramsey interferometric scheme relies upon nonlinearity only for generating squeezing (in step (b-c)) and there is no critical dependence on the exact knowledge of the total atomic number unlike in Sec.~IIIA. The situation is similar to that explored in some recent squeezed Ramsey interferometric experiments, for example Ref.~\cite{Strobel-2015} clearly demonstrated phase resolution below standard quantum limit even with $N \approx 380 \pm 15$, a relatively small atomic number and large relative fluctuation of $N$.

\section{Sensitivity to rotation}

The dynamics of cold atoms in a ring lattice and the physical pictures used to describe it can be characterized by the sensitivity to rotation. Here we provide a very basic order of magnitude comparative numerical estimate using a simple criterion of the minimum resolution $\Delta\omega$ about $\omega=0$
set by the averaged standard deviation
\bn |\langle 2J_z(0)\rangle -\langle 2J_z(\Delta\omega)\rangle|\geq \sigma_z(0)+\sigma_z(\Delta\omega)\label{rotsens},\en
where $2J_z(\omega)$ is the mean differential modal population at the time of measurement $t_s$.
Our simulations assume $N=10^{4}$ sodium atoms with $m=3.8\times 10^{-26}$ kg, ring radius $R= 10^{-5}$ m, mode $q= 5$, and radial confinement $\omega_r=2\pi\times 100$ Hz. These set our units $\epsilon=2.9\times10^{-33}$ J and $\tau=0.23$ s and $\omega_0=28$ rad/s.

In the non-interacting linear limit with $\chi=0$ in Eq.~\eqref{Hama}, the complementary probabilities of the two modes indicate a binomial distribution, with each of the $N$ atoms constituting an independent trial, so that $\sigma_z(\omega)={\textstyle \frac{1}{\sqrt{N}}}|a(\omega)||b(\omega)|$. For lattice depth  $u=\epsilon \pi/400$, maximum swap time is $t_s=200 \tau=46$ s, and the criterion in Eq.~\eqref{rotsens} yields $\Delta \omega=\omega_0 u/(2q\sqrt{N})=2.2\times 10^{-4}$ rad/s.  In the corresponding nonlinear case, the interaction strength is $N\chi=20.6\omega_0$. Using Fig.~\ref{Figure-7}(c), where $u=5.25 \omega_0 > u_c=N\chi/4$,the critical value, we estimate sensitivity  of $\Delta \omega=1.5\times 10^{-3}\omega_0=4.1\times 10^{-2}$ rad/s.

For a uniform comparison, we normalize by the duration, in the linear case, $\Delta\omega\times \sqrt{t_s}=1.5\times 10^{-3}$ rad/s/$\sqrt{\rm Hz}$ and for the nonlinear case $\Delta\omega\times \sqrt{t_s}=1.5\times 10^{-2}$ rad/s/$\sqrt{\rm Hz}$ with $t_s=0.57\tau=0.13$ s. Despite the advantage of a much shorter cycle duration the nonlinear case remains an order of magnitude less sensitive to rotation near the critical point, underscoring that fluctuations more than erase the nonlinear advantage indicated by mean field dynamics alone.

Squeezing, however, can reverse that situation. For the Ramsey interferometric scheme the evolution of the standard deviation $\sigma_z$ listed in Table~I, shows that squeezing along the quadrature of interest $J_z$ leads to a significant increase in sensitivity over the Poisson value:  factor of 5 and 10 reduction in fluctuation for $N$=1,000 and $N$ =10,000 respectively.
Shorter cycle duration (from Table I) causes time-normalized sensitivity to increase by factors of about 30 and 120. Using our estimate for the linear case, the enhanced rotation sensitivity would be $\Delta\omega \sim 2 \times 10^{-5}$ rad/s or $\Delta\omega\times \sqrt{t_s}\sim 10^{-5}$ rad/s/$\sqrt{\rm Hz}$.

Obvious scaling with number of particles indicates that with $N\geq 10^5$ further improvement in sensitivity is possible. The single cycle sensitivity can be enhanced by increasing the sensing time $ t_{\rm sensing}$, but can compromise the time-normalized sensitivity.  Since the $\langle \op{J}_z\rangle$ dependence on $\omega$ is linear, the sense of rotation can be resolved.

\section{Discussion and Conclusion}
Ring-shaped lattices discussed here can be created using two Laguerre-Gaussian beams \cite{Cataliotti-ring-lattice, padgett} with opposite orbital angular momentum $LG_{\pm}\propto \sqrt{I_0}\exp(\pm iq\phi)$. Combining these beams coherently with different complex amplitudes $a_{\pm}$, such that $|a_+|^2+|a_-|^2=1$, one can achieve the required intensity structure.  The Raman techniques of transferring quantized orbital angular momentum (OAM) from light beams to condensates \cite{KC-Wright}, allow one to create circulating condensate with the wavefunction of the form we use.

To detect values of $\hat J_z$, the procedure can be reversed: a Raman process transforms atoms circulating in the $a_+$ mode  to a nonrotating BEC and then the proportion of the nonrotating atoms is measured.  In \cite{ramanathan}, such measurement was performed by releasing the condensate and observing the size of the central hole in the interference pattern produced by the free-falling atoms. Alternately, as discussed in  \cite{OKD2015}, one can measure $\hat J_z$ by coupling the ring resonator to a linear atomic waveguide (formed, e.g., by a red-detuned horizontal laser beam) positioned tangentially near the ring. Atoms circulating in the opposite orientations would leak to the waveguide and propagate in opposite directions towards the waveguide ends where they can be detected.

The physical parameters used in our simulations correspond to scenarios already realized in the context of toroidal traps \cite{ramanathan}, the duration $\sim 40$ s (and longer since) of persistent currents observed in such experiments accommodate our estimates.  The demonstration of the nonlinear critical behavior as well as the spin squeezing effects described here therefore should be accessible within the spectrum of current experimental capabilities.

Going beyond proof of principle, to utilize the nonlinear behavior for rotation sensing will certainly require more effort, and perhaps the biggest challenge would be to determine the number of particles in the system with the required level of precision. In this regard, there have been some remarkable developments in recent years where individual atoms in lattices could be imaged \cite{Greiner}; subsequent melting of such a lattice can be used to initiate experiments with well-defined number of atoms.

Our estimates of optimal rotation sensitivity of $\Delta\omega\sim 10^{-5}$ rad/s or $\Delta\omega\times \sqrt{t_s}\sim 10^{-5}$ rad/s/$\sqrt{\rm Hz}$ for the parameters used approaches but falls short of the current state of the art capabilities of atom interferometers at the order of $10^{-7}$ rad/s/$\sqrt{\rm Hz}$ \cite{kasevich-2011} $10^{-7}$ rad/s \cite{Kasevich-2015} respectively. However, a relevant comparison has to take into account that we use substantially less atoms, and notably the area of our ring at $\sim 10^{-10}$ m$^2$ is several orders of magnitude smaller than corresponding parameters in such experiments, so that normalization by possible parallel realizations can make our approach competitive. While there will certainly be challenges to overcome in implementation, that has to be placed in the context that currently used methods are the result of almost two decades of engineering and experiments.

Aside from the possible applications for rotation sensing, our study has shown that toroidal lattices offer a novel system for studying spin-squeezing and nonlinear dynamical effects like self-trapping, with the option of easily including non-inertial effects. Of particular relevance for future studies, we have shown that mean field theory is inadequate for accurate analysis of such a system and quantum fluctuations and correlations have to be taken into consideration.

\acknowledgments
M.K. and T.O. acknowledge the support of Czech Science Foundation Grant No. P205/10/1657. K.K.D. acknowledges support from the National Science Foundation under Grants No. PHY-1313871 and PHY11-25915 and of a PASSHE-FPDC grant and a research grant from Kutztown University.

\appendix

\section{Moments in Gaussian approximation}
\label{AppMoments}

In this appendix we present the equations of motion, used in Sec. IV, for the first and second moments of the operators $\hat{J}_{i}, (i=x,y,z)$, defined by
\begin{eqnarray}
{\cal J}_s &\equiv& \langle \hat{J}_s \rangle, \\
V_{nl} &\equiv& \frac{1}{2}\left\langle (\hat{J}_n-{\cal J}_n)(\hat{J}_l-{\cal J}_l)  +(\hat{J}_l-{\cal J}_l) (\hat{J}_n-{\cal J}_n)\right\rangle.
\n\end{eqnarray}
They follow from the Heisenberg equations $i\hbar \langle \dot{A}\rangle=\langle [A,H]\rangle $ on using the approximation $\langle J_k J_s J_p\rangle \approx {\cal J}_k {\cal J}_s{\cal J}_p
+ {\cal J}_k V_{sp} + {\cal J}_s V_{pk} + {\cal J}_p V_{ks}$ valid for Gaussian states. Thus we find
\begin{eqnarray}
\label{eqJx}
\dot{{\cal J}}_x &=& u_y {\cal J}_z + 2q\omega {\cal J}_y + 2\chi
\left( {\cal J}_y{\cal J}_z + V_{yz}\right) , \n\\
\dot{{\cal J}}_y &=& -u_x {\cal J}_z - 2q\omega {\cal J}_x - 2\chi
\left( {\cal J}_x{\cal J}_z + V_{xz}\right) , \n\\
\dot{{\cal J}}_z &=& u_x {\cal J}_y-u_y {\cal J}_x ,
\end{eqnarray}
and
\begin{eqnarray}
\dot{V}_{xx} &=& 2u_y V_{xz}+4q\omega V_{xy} + 4\chi ({\cal J}_z V_{xy}
 + {\cal J}_y V_{xz}) , \n\\
\dot{V}_{yy} &=& -2u_x V_{yz}-4q\omega V_{xy} - 4\chi ({\cal J}_z V_{xy}
 + {\cal J}_x V_{yz}) ,\n\\
\dot{V}_{zz} &=& 2u_x V_{yz} - 2u_y V_{xz}\n\\
\dot{V}_{xy} &=& -u_x V_{xz} + u_y V_{yz} - 2q\omega (V_{xx}-V_{yy}) \nonumber \n\\
& & -2\chi \left[ {\cal J}_z (V_{xx}-V_{yy}) + {\cal J}_x V_{xz} -  {\cal J}_y V_{yz} \right] ,\n\\
\dot{V}_{xz} &=& u_x  V_{xy} +  u_y (V_{zz}-V_{xx}) +2q\omega V_{yz} \nonumber \n\\
& & +2\chi \left( {\cal J}_z V_{zy} + {\cal J}_y V_{zz}  \right),\n\\
\dot{V}_{yz} &=&  u_x (V_{yy}-V_{zz}) - u_y V_{xy} - 2q\omega V_{xz} \nonumber \n\\
& & - 2\chi \left( {\cal J}_z V_{xz} + {\cal J}_x V_{zz}  \right).
\label{eqVyz}
\end{eqnarray}

\end{document}